\input{aipcheck}

\documentclass[
    ,final            
  ]
  {aipproc}

\layoutstyle{8x11double}

\def\degree{\hbox{$^\circ$}}
\newcommand{\rxte}{\textit{RXTE}}
\newcommand{\sax}{\textit{BSAX}}
\newcommand{\bsax}{{\textit{Beppo}SAX}}
\newcommand{\ginga}{\textit{Ginga}}
\newcommand{\fouruoh}{\mbox{4U~0115$+$63}}
\newcommand{\mxoh}{\mbox{MX~0656$-$07}}
\newcommand{\vela}{\mbox{Vela~X$-$1}}
\newcommand{\gx}{\mbox{GX~301$-$2}}
\newcommand{\wray}{Wray\,977}
\newcommand{\xray}	{\mbox{X-ray}}
\newcommand{\xrays}	{\mbox{X-rays}}
\newcommand{\ltsim}{\lower.5ex\hbox{$\; \buildrel < \over \sim \;$}}
\newcommand{\gtsim}{\lower.5ex\hbox{$\; \buildrel > \over \sim \;$}}

\newcommand{\aprx}	{\mbox{$\sim$}}

\begin{document}

\title{Timing and Spectroscopy
of Accreting X-ray Pulsars: the State of Cyclotron Line Studies}

\author{W.A. Heindl}{
  address={Center for Astrophysics and Space Sciences, University of
  California, San Diego}
}

\author{R.E. Rothschild}{
  address={Center for Astrophysics and Space Sciences, University of
  California, San Diego}
}



\author{W. Coburn }{
 address={Space Sciences Laboratory, University of  California, Berkeley}
}

\author{R. Staubert}{
 address={Institut f\"ur Astronomie und Astrophysik, T\"ubingen}
}

\author{J. Wilms }{
 address={Institut f\"ur Astronomie und Astrophysik, T\"ubingen}
}

\author{I. Kreykenbohm}{
  address={INTEGRAL Science Data Center}
  ,altaddress={Institut f\"ur Astronomie und Astrophysik, T\"ubingen}
}

\author{P. Kretschmar}{
  address={INTEGRAL Science Data Center}
  ,altaddress={Max-Planck-Institut f\"ur Extraterrestrische Physik }
}

\begin{abstract}
 A great deal of emphasis on timing in the \rxte\ era has been on pushing
toward higher and higher frequency phenomena, particularly kHz
QPOs. However, the large areas of the \rxte\ pointed instruments provide
another capability which is key for the understanding of accreting
\xray\ pulsars -- the ability to accumulate high quality spectra in a
limited observing time.  For the accreting \xray\ pulsars, with their
relatively modest spin frequencies, this translates into an ability to
study broad band spectra as a function of pulse phase. This is a
critical tool, as pulsar spectra are strong functions of the geometry
of the ``accretion mound'' and the observers' viewing angle to the
\aprx$10^{12}$\,G magnetic field.  In particular, the appearance of ``cyclotron
lines'' is sensitively dependent on the viewing geometry, which must
change with the rotation of the star.  These spectral features, seen
in only a handful of objects, are quite important, as they give us our
only direct measure of neutron star magnetic fields. Furthermore, they
carry a great deal of information as to the geometry and physical
conditions in the accretion mound. In this paper, we review the status
of cyclotron line studies with the
\rxte. We present an overview of phase-averaged results and give
examples of observations which illustrate the power of phase-resolved
spectroscopy.

\end{abstract}

\maketitle


\section{Introduction}

In this work, we review the status of \rxte\ cyclotron line studies 
in the classical accreting \xray\ pulsars and discuss the
observational requirements needed to further our understanding in this
area.  We specifically do not consider the new class of
\emph{millisecond} accreting pulsars, as they are the subject of other
papers in this proceedings.

Cyclotron lines, or more precisely ``cyclotron resonance scattering
features'' (CRSFs) are formed at or near the neutron star magnetic
polar cap where electron motions perpendicular to the field are
quantized in Landau orbits.  This gives rise to increased magnetic
Compton scattering opacity at the (harmonically spaced) Landau
energies resulting in absorption-line-like features in the emergent
spectra. Owing to the \aprx10$^{12}$\,G field strength, the Landau
transitions have energies in the hard \xray\ band: $E_{\rm n,s} = (n +
\frac{1}{2} + s) E_{\rm cyc}$, where $n$ is the principle quantum
number, $s=\pm 1/2$ is the electron spin, and $E_{\rm cyc} =
\frac{\hbar e B}{m_e} = 11.6 \frac{B}{10^{12}\,\rm G}$\,keV.   
Thus, the cyclotron line energy gives a direct measure of the magnetic
field in the scattering region.  Of course, the observed line energy
will be redshifted in the strong gravitational field of the neutron
star such that $B_{12} = (1 + z) \frac{E_{\rm obs}}{11.6\rm \,keV}$,
where $B_{12}$ is the field strength in $10^{12}$\,G and $z$ is the
redshift in the scattering region.  In addition, when relativistic
effects are considered \cite{Harding1991} the line spacing is no longer
expected to be harmonic: 
$E_{\rm cyc} \propto [( 1 + 2n\frac{B}{B_{\rm crit}}{\rm sin}^2\theta )^{1/2} - 1 ]/{\rm sin}^2\theta$
where $\theta$ is viewing angle with respect to the magnetic field
and ${B_{\rm crit}} = m^2c^3/e\hbar = 4.4\times10^{13}\,\rm G$.  For
magnetic fields in the $10^{12}$\,G range, these shifts are
\ltsim10\%.

\begin{figure}
\includegraphics[width=0.4\textwidth]{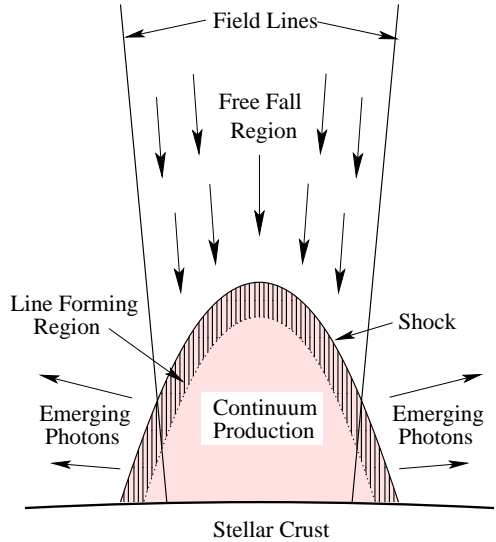}
  \caption{Schematic diagram of the ``accretion mound'' showing the
  line-forming region as a discrete layer covering the continuum
  production zone. \label{fig:mound}}
\end{figure}

\begin{figure}
\includegraphics[width=0.4\textwidth]{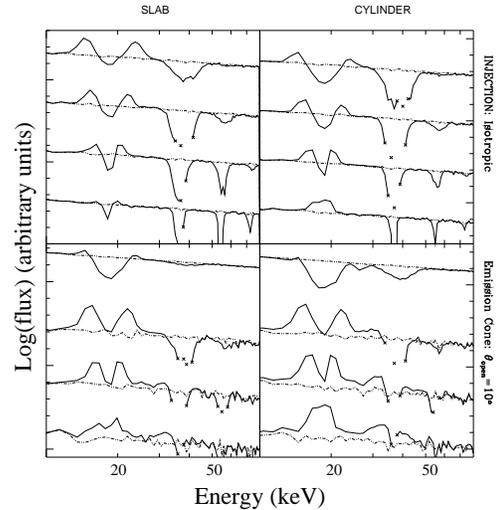}
  \caption{From \cite{Araya2000}. Monte Carlo models of emergent
  photon spectra as a function of viewing angle, emission mound
  geometry, and continuum injection geometry.  Each panel shows the
  injected (dashed line) and detected (solid line) spectra for four
  viewing angles ranging (top to bottom) from nearly parallel to
  nearly perpendicular to the magnetic field.  The left panels are for
  a slab geometry accretion mound and the right panels are a
  cylindrical mound.  Finally the top two panels are for isotropic
  continuum injection while the the bottom panels are for conical
  injection about the magnetic field direction.  \label{fig:monte}}
\end{figure}

These considerations lay down the basic theoretical principles for
cyclotron line formation.  However, the practical reality is that the
situation is quite complex.  Monte Carlo simulations by several
authors (see e.g., \cite{Araya1999,Araya2000,Isenberg1998}) show that
when the physical geometry of the emitting region and the observing
aspect are considered, complex line shapes, particularly for the
fundamental, can be produced.  Figure~\ref{fig:mound} is a schematic
of the model geometry assumed by these Monte Carlo studies.  The
\xrays\ are produced in an ``accretion mound''.  This consists of an
underlying continuum producing volume above which photons must
propagate through a line forming region characterized by a magnetic
field strength, electron temperature, and Thomson optical depth.  In
various models, the electron temperature, the geometry of the mound,
the angle of the magnetic field with respect to the mound, and the
injected continuum beam pattern can be varied. Figure~\ref{fig:monte},
(from \cite{Araya2000}) compares predicted line shapes for various
viewing angles for both slab and cylindrical mound geometries when
continuum photons are injected isotropically and in a cone.  In these
models, the fundamental line only appears as a simple absorption
feature in the case of a slab-shaped mound with isotropic injection
viewed nearly perpendicular to the magnetic field axis.  In some
cases, the fundamental is flanked by apparent emission wings, and can
even appear as an \emph{emission} line.  Furthermore, it is quite clear that the
expected line shape is a strong function of viewing geometry.  And, it
is precisely this viewing angle which is changing as the neutron star
rotates.  Thus, it would be quite naive to expect to make a clean
measurement of a cyclotron line without performing phase-resolved
spectroscopy.  Conversely, if the phase average spectrum does appear
simple, then we can place limits on the variation of the viewing angle
through the pulse.

\section{\rxte\ phase-average results}

In this section, we give an overview of measurements of cyclotron
lines using the \rxte.  Table~\ref{tab:lines} lists all of the pulsars
where cyclotron lines have been convincingly detected.  The instrument
which was used in the discovery of the line is given as well as
whether the line has been detected with \rxte.  A total of 14 pulsars
have secure line detections, and all but two (V~0332$+$53 and
A~0535$+$26: transients which have been inactive during the \rxte\
lifetime) have been measured.  Furthermore, five of the lines have
been discovered with \rxte\ or with both \bsax\ and \rxte.

\begin{table}
\caption{List of pulsars with securely detected cyclotron lines.  The
discovery instrument is listed along with the discovery reference
and whether the line has been observed with \rxte. \label{tab:lines}}
\begin{tabular}{lllc} 
\hline
\tablehead{1}{l}{b}{Source\\}
  & \tablehead{1}{l}{b}{Energy\\(keV)}
  & \tablehead{1}{l}{b}{Discovery\\Instrument}
  & \tablehead{1}{l}{b}{\rxte?\\} \\
\hline
4U~0115$+$63$^\dagger$   	& 12 \cite{Wheaton79}    &  \emph{HEAO-1} & Y \\
4U~1907$+$09$^{\dagger\ddag}$  & 18 \cite{Makishima92}    &  \emph{Ginga}  & Y \\
4U~1538$-$52$^{\ddag}$   	& 20 \cite{Clark90}    &  \emph{Ginga}  & Y \\
Vela~X$-$1$^{\dagger\ddag}$    & 25 \cite{Kendziorra1992}    &  HEXE          & Y \\
V~0332$+$53             	& 27   \cite{Makishima90}   &  \emph{Ginga}  & N \\
Cep X$-$4            	& 28 \cite{Mihara91}   &  \emph{Ginga}    & Y \\
Cen~X$-$3$^{\ddag}$    	& 28.5 \cite{Santangelo98,Heindl99a}  &  \emph{RXTE}/\sax & Y \\
X~Per         	& 29  \cite{Coburn01a}  &  \emph{RXTE}   & Y \\
XTE~J1946$+$274         & 36 \cite{Heindl01}   &  \emph{RXTE}/\sax & Y \\
MX0656$-$072           	& 36  \cite{Heindl2003} &  \emph{RXTE}   & Y \\
4U~1626$-$67            & 37 \cite{Heindl99a,Orlandini98A}  &  \emph{RXTE}/\sax & Y \\
GX~301$-$2$^{\ddag}$      	& 37 \cite{Makishima92}   &  \emph{Ginga}  & Y \\
Her~X$-$1$^{\ddag}$       	& 41 \cite{Truemper78}  &  Balloon       & Y \\
A~0535$+$26               	& 50?, 110 \cite{Kendziorra94A}  &  HEXE    & N \\
\hline\hline
\multicolumn{4}{l}{ $^\dagger$ objects with $>$ 1 harmonic observed} \\
\multicolumn{4}{l}{ $^\ddag$ high inclination system}
\end{tabular}
\end{table}

\subsection{MX~0656-072: A new line in an old
transient \label{sec:mxoh} }

\begin{figure}
\includegraphics[width=0.45\textwidth,bb=123 134 530 525,clip]{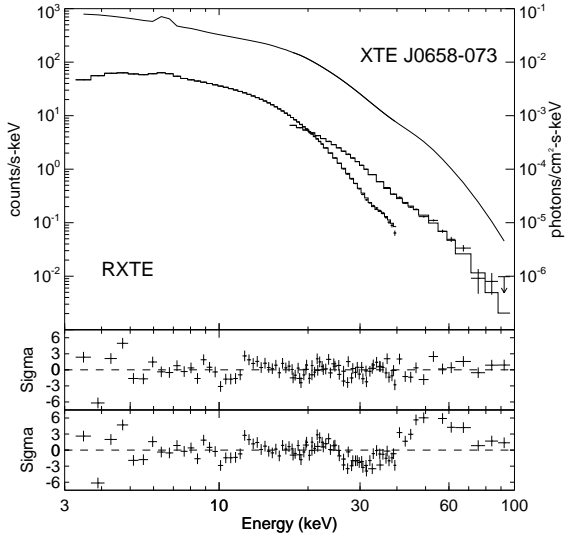}
  \caption{\label{fig:mxoh} The phase average spectrum of \mxoh\ observed with
  \rxte. The middle panel shows residuals for the full model including
  a cyclotron line, while the bottom panel is the best fit continuum
  with no cyclotron line.  The line is apparent as the dip at
  \aprx30\,keV and the underestimation of the continuum above 40\,keV.}
\end{figure}

As an example of a cyclotron line seen in a phase averaged spectrum,
we show the recent discovery of the line in \mxoh\
(=XTE~J0658$-$072). \mxoh\ was discovered with SAS-3 in October 1975
\cite{Clark1975} and was seen again with Ariel-V in March 1976
\cite{Kaluzienski1976}.  At the time, it was not recognized as a pulsar.  
The source became active again only in 2003 October
\cite{Remillard2003}, and an \rxte\ observation on 2003 October 19-20
revealed that it is a 160\,s accreting pulsar \cite{Morgan2003}.  Its
optical counterpart is a Be star \cite{Pakull2003}, making it one of the
transient Be star/\xray\ pulsar binaries.  Analyzing the phase average
spectrum from this observation, we discovered a cyclotron line at
\aprx36\,keV \cite{Heindl2003}.  The spectrum is shown in Figure~\ref{fig:mxoh}.  
We fit the continuum with a power law which breaks smoothly to a power
law times an exponential cutoff at high energies (the ``modified power
law cutoff model'', MPLCUT \cite{Coburn2002}).  An iron line was also
required. The cyclotron line is modeled by an absorption line with a
Gaussian optical depth profile (see \cite{Coburn2002} for the exact
functional form). The best-fit line parameters are: $\rm E_{cyc} =
36\pm1$\,keV, $\rm \sigma_{cyc} = 7.5\pm1.0$\,keV, $\rm \tau_{cyc} =
0.33\pm0.05$.  $\rm \sigma_{cyc}$ and $\rm
\tau_{cyc} $ are the width and maximum of the optical depth profile.


\subsection{Correlations of Spectral Parameters}

\begin{figure}
\includegraphics[width=0.4\textwidth]{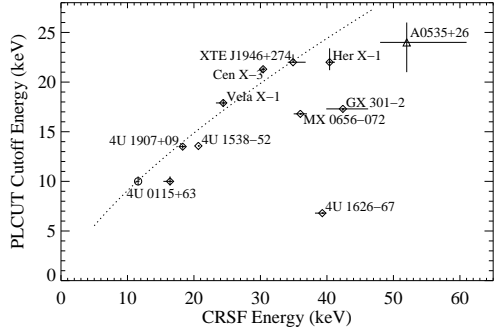}
\caption{\label{fig:hecutvse} Continuum break energy plotted versus
cyclotron line energy. After Figure~9 of
\cite{Coburn2002}. Hashed regions indicate phase space where the
\rxte\ is not sensitive to lines.  Except for A~0535$+$26, all data
points are from \rxte.}
\end{figure}

\begin{figure}
\includegraphics[width=0.4\textwidth]{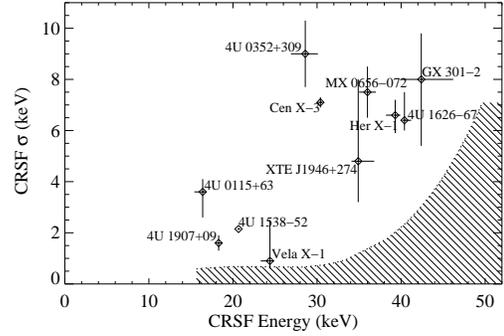}
  \caption{\label{fig:sigmavse} Cyclotron line width plotted versus
cyclotron line energy. The width is the sigma of the Gaussian optical
depth profile. After Figure~7 of \cite{Coburn2002}. Hashed regions indicate phase space where the
\rxte\ is not sensitive to lines.}
\end{figure}

\begin{figure}
\includegraphics[width=0.4\textwidth]{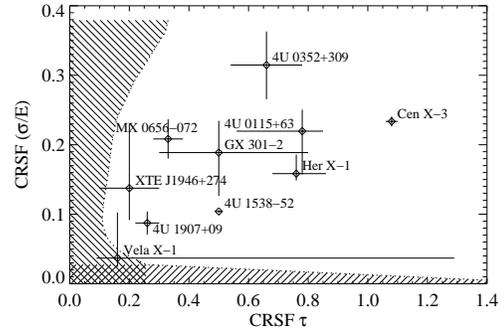}
  \caption{\label{fig:ratiovsdepth} Fractional cyclotron line width
plotted versus cyclotron line energy. The fractional width is the sigma of the
Gaussian optical depth profile divided by the line energy. After Figure~8 of
\cite{Coburn2002}.  Hashed regions indicate phase space where the
\rxte\ is not sensitive to lines.}
\end{figure}

Coburn \cite{Coburn2001} and Coburn \textit{et al.}
\cite{Coburn2002} examined the phase average spectra of all the
pulsars observed with high statistics with \rxte.  In the cases where
no cyclotron line was found (12 objects), the authors placed upper
limits on the width and optical depth of any line as a function of
energy.  In order to look for class-wide relationships among the 12
cyclotron line pulsars, they fit their spectra to the common spectral
model (MPLCUT plus iron emission) including a cyclotron line as
described above.  They then searched for correlations between model
parameters.  Three significant correlations were noted: 1. between the
cyclotron line energy and the continuum break energy, 2. between the
width of the cyclotron line and its energy, and 3. between the
fractional width of the cyclotron line and its depth.
Figures~\ref{fig:hecutvse}--\ref{fig:ratiovsdepth} are plots of the
fit parameters showing these three correlations.  In these plots, the
hashed regions indicate where \rxte\ is not sensitive to lines for
typical source brightnesses and moderate observing times.  The
fact that the observed points do not fill the available phase space
shows that the correlations are not selection effects.  The first two
of these correlations have been noted previously (1:
\cite{Makishima92}, \cite{Makishima99}; 2: \cite{dalFiume00},
\cite{Heindl00A}), while the third is new.  Importantly, this is the 
first time that these correlations have been demonstrated in a uniform
analysis of data from a single set of instruments.  

The correlation between cyclotron line energy and continuum break
energy was originally noted by \cite{Makishima92} and
\cite{Makishima99} who compared fit parameters from different
missions.  With the exception of A~0535$+$26 (plotted at the lower of
the two possible fundamental line energies \cite{Araya1996}), all the
data in Figure~\ref{fig:hecutvse} come from \rxte.  Below \aprx35\,keV,
the correlation appears clean.  However, at higher line energies the
case is not so clear.  We discount 4U~1626$-$67, as its break energy
is highly variable with pulse phase and so the phase average spectrum
is a poor representation.  However, this still leaves \gx\ and \mxoh\
in poor agreement with the correlation. One possibility is that these
sources are like \vela.  In \vela, the strong observed line near
50\,keV is in fact a harmonic, and phase resolved spectroscopy has
confirmed (see below) that the fundamental is weak, lying near 25\,keV.
Moving the line energies down by a factor of two would bring \gx\ and
\mxoh\ in good agreement with the correlation.  Another, perhaps more
likely explanation is that, like 4U~1626$-$67, the phase average
spectrum is misleading for these sources. Finally, of course, the
correlation may simply break down at these energies.  In any case, the
correlation suggests that the continuum cutoff is a magnetic effect
and that it is not the temperature of the accretion mound alone that
determines the hardness of high energy spectrum.

To first order, a correlation between the line width and energy is
expected. If the line width ($\Gamma_{cyc}$) is dominated by the
temperature of the electrons ($kT_e$), then according to
\cite{Meszaros1985}, $\Gamma_{cyc} \sim E_{cyc} \left(
\frac{8ln(2)kT_e}{m_ec^2}
\right)^{\frac{1}{2}} \left| cos\theta \right| $.  The fact that this
correlation is observed, however, implies that there is neither a
large variation in electron temperature nor in viewing angle,
$\theta$, from source to source.  If there were, then the correlation
should be destroyed.  The stricture on $cos\theta$ is quite
interesting given the prevalence of high inclination systems among the
cyclotron line pulsars (see Table~\ref{tab:lines}). This suggests that
the magnetic and spin axes in these systems tend to be nearly
aligned, and that either they are born that way or that accretion
tends to align them.

The correlation between the fractional line width and energy is new
and in fact quite surprising.  This correlation says that broad lines
tend to be deep and narrow lines tend to be shallow.  However, both
from simple considerations of the magnetic scattering cross sections
at finite temperatures \cite{Araya1999} and from Monte Carlo models
(see Figure~\ref{fig:monte}), it is the narrow lines that should be
deep.  

\section{Selected \rxte\ phase-resolved results}

In this section, we discuss three pulsars where we have carried out
pulse phase resolved spectroscopy.  These observations show the power
of the large collecting areas or the \rxte\ instruments to reveal
detailed variations in the spectrum and hint at possibilities for a
future mission that will perform such observations of much weaker pulsars.

\subsection{\fouruoh}

\begin{figure}
\includegraphics[width=0.4\textwidth]{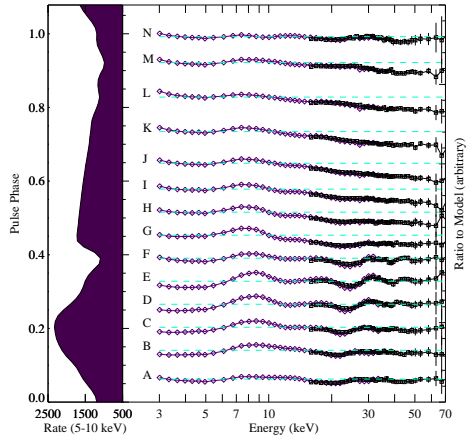}
  \caption{\label{fig:0115prs} Normalized ratios of the \rxte\ data in
  14 phase bins to a smooth
  continuum model.  On the left, the corresponding pulse profile is
  plotted vertically.  The dips in the ratios indicate the presence of
  multiple cyclotron line harmonics.}
\end{figure}

The Be/\xray\ binary transient pulsar \fouruoh\ was one of the
earliest cyclotron line pulsars and was the first to show both a
fundamental and a harmonic line \cite{White1983}.  It is still the
pulsar with the lowest fundamental energy, making it the best
candidate for multiple harmonics to be observed.  In 1999 March-April,
\fouruoh\ underwent an outburst that was observed both with \rxte\
\cite{Heindl1999} and \bsax\,\cite{Santangelo1999ApJ}.  
The results from the two observatories are in good agreement.  We
present the \rxte\ results here.  Figure~\ref{fig:0115prs} shows the
spectrum as a function of pulse phase.  At each of 14 phases
corresponding to the pulse profile (plotted vertically), the PCA and
HEXTE spectra are plotted, divided by a smooth continuum model and
normalized to unit intensity.  This reveals strong variability in the
spectrum, both in overall hardness as well as in the presence of
cyclotron lines.  The wiggles (e.g., phase D) bely the presence of
up to five cyclotron line harmonics.  This is in fact the only source
where more than two lines have been seen.

\begin{figure}
\includegraphics[width=0.4\textwidth, bb=124 157 530 518]{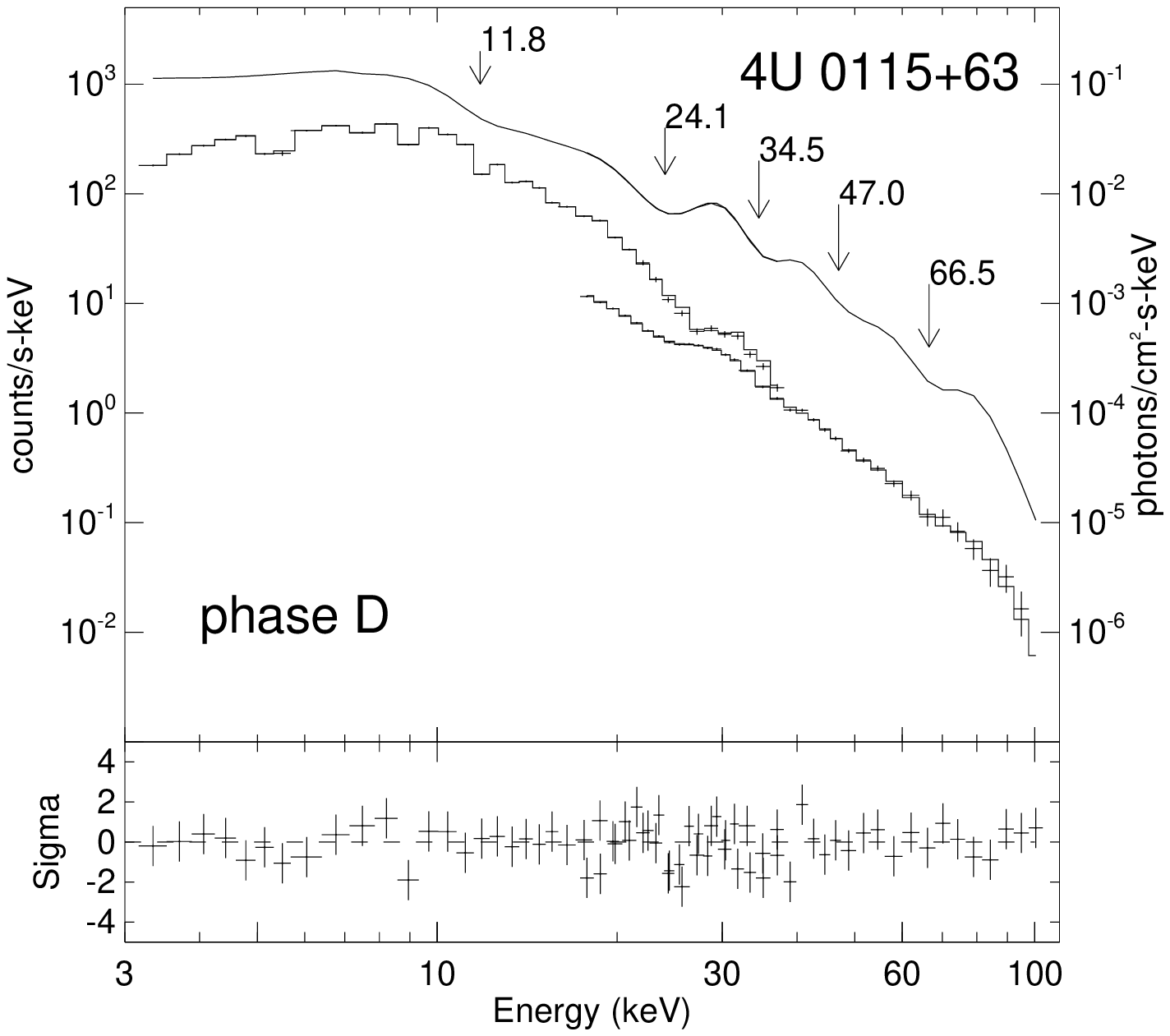}
  \caption{\label{fig:0115D}}
\end{figure}

Detailed fitting of these spectra reveals that as many as five lines
(the fundamental plus four harmonics) are detected.
Figure~\ref{fig:0115D} shows a fit to the phase D spectrum, with the
required lines indicated.  The fundamental has been fit with a complex
shape, and its energy is hard to determine.  As expected from theory,
however, the rest of the lines follow a harmonic relationship with a
spacing of half of the energy of the \aprx24\,keV harmonic.



\begin{figure}
\includegraphics[width=0.4\textwidth, bb=100 84 523 523]{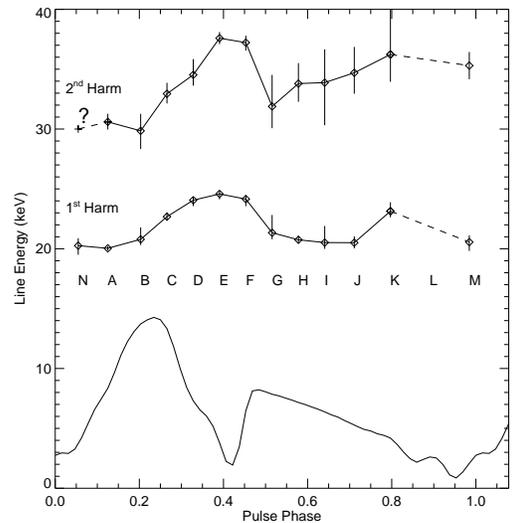}
  \caption{\label{fig:evsph} The energies of the \fouruoh\ first and
  second harmonic lines as a function of pulse phase plotted above the
  arbitrarily normalized pulse profile.  The harmonic relation of the
  line energies is maintained within statistics at all phases where
  the lines are detected.}
\end{figure}

Finally, figure~\ref{fig:evsph} shows the evolution of the first and
second harmonic line energies with pulse phase.  It is clear
that we observe regions with a 20\% range in magnetic field. This
emphasizes the need for phase resolved spectroscopy as these
variations are integrated in a phase average spectrum.  Such
variations could be due, for example, to quadrupole and higher field
components or an offset of the dipole with respect to the center of
the star.

\subsection{\vela}

\vela\ is a 283\,s pulsar in an 8.964\,d orbit with a B0.5Ib
supergiant.  The radius of the orbit is only \aprx1.7 times the B star
radius, and with this small separation, the neutron star accretes from
the intense stellar wind.  Observations with HEXE on \textit{Mir}
\cite{Kendziorra1992}, \ginga\ \cite{Mihara1995}, 
\rxte\ \cite{Kreykenbohm1999}, and \bsax\
\cite{Orlandini1998} all show evidence of a cyclotron line at
\aprx50\,keV.  There have also been reports with 
\ginga\ \cite{Mihara1995} and HEXE \cite{Kendziorra1992}
of a line near 24\,keV.  This line would be the fundamental, making
the \aprx50\,keV an harmonic.  Observations with \bsax\ have not
confirmed this line. However, phase resolved observations with \rxte\
now find that there is a weak, phase variable line near 25\,keV.

\begin{figure}
\centerline{\includegraphics[width=0.4\textwidth]{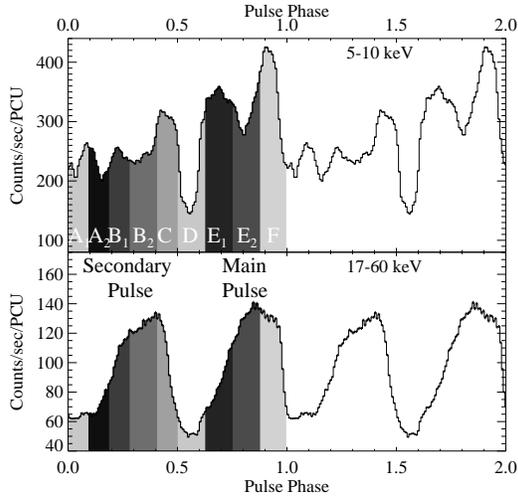}}
  \caption{\label{fig:velapuls}From \cite{Kreykenbohm2002}. The \vela\ pulse
  profile in two energy bands.  The shaded regions indicate phase bins
  used for phase resolved spectroscopy (see figure~\ref{fig:velacrsfs}).}
\end{figure}

\begin{figure}
\centerline{\includegraphics[width=0.4\textwidth]{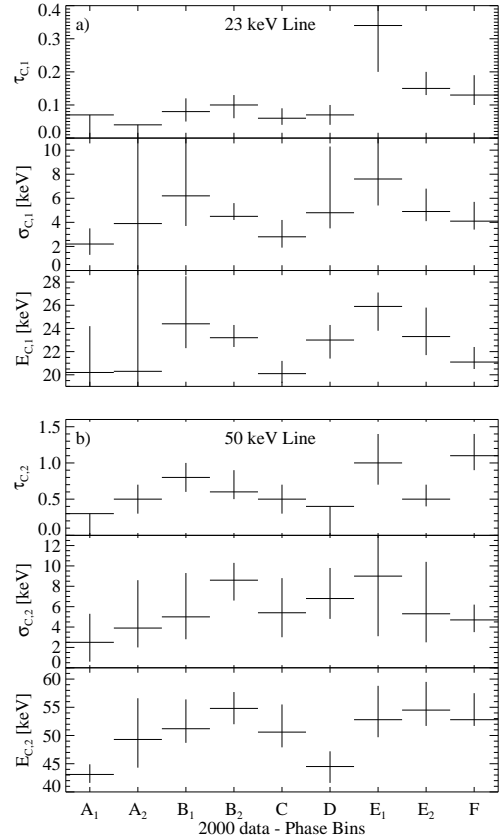}}
\caption{\label{fig:velacrsfs} From \cite{Kreykenbohm2002}. Variations of the
\vela\ fundamental and harmonic cyclotron line parameters with pulse
phase for the phase bins defined in figure~\ref{fig:velapuls}.}
\end{figure}

Figure~\ref{fig:velapuls} from \cite{Kreykenbohm2002} shows the \vela\ pulse
profile in two energy bands from a 2000 January \rxte\ observation.
The profile has two, simple main peaks at high energies that are
subdivided into multiple peaks at low energies.
The phase ranges for phase resolved spectroscopy are indicated. The
evolution of cyclotron line parameters, for both the fundamental and
the harmonic, is shown in figure~\ref{fig:velacrsfs}.  Note that, as
indicated by the optical depth, the fundamental is not significantly
detected over much of the secondary pulse.  Indeed, this phase
variability of the line is strong evidence that it is real and not an
artifact of the spectral fitting.  It is likely due to the combination
of this pulse phase variability and the strongly time variable
photoelectric absorption by the stellar wind (N$_H$ can be well over
10$^{23}$\,cm$^{-1}$) that it has been difficult to confirm this line.

\subsection{\gx}

\begin{figure}
\centerline{\includegraphics[width=0.45\textwidth]{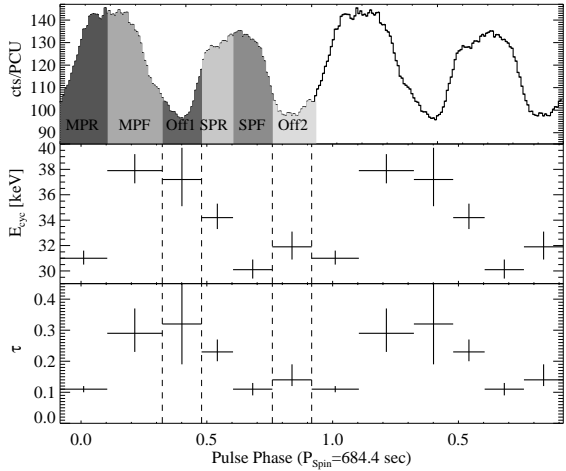}}
\caption{\label{fig:gxprs} From \cite{Kreykenbohm2004}.  The \gx\ pulse
profile and phase resolved spectral fits of the cyclotron line energy
and depth.}
\end{figure}

The neutron star in \gx\ orbits the early type B-emission line star
\wray.  The 41.5\,d orbit is eccentric ($e=0.462$), and just prior to
periastron passage the neutron star intercepts the gas stream from
\wray\ \cite{leahy91a,leahy02a}, resulting in an extended \xray\
flare.  In 2000 October, Kreykenbohm et al. \cite{Kreykenbohm2004} made a
long observation with \rxte\ spanning this pre-periastron flare.
Figure~\ref{fig:gxprs} shows the resulting pulse profile along with
the phase behavior of the cyclotron line parameters. Note the
similarity of the pulse to \vela. Both sources show a double-peaked
profile with the two peaks having similar strength and shape.  Again,
the line energy is strongly variable, shifting by more than 20\%.

\begin{figure}
\begin{minipage}{0.45\textwidth}
\centerline{\includegraphics[width=0.9\textwidth]{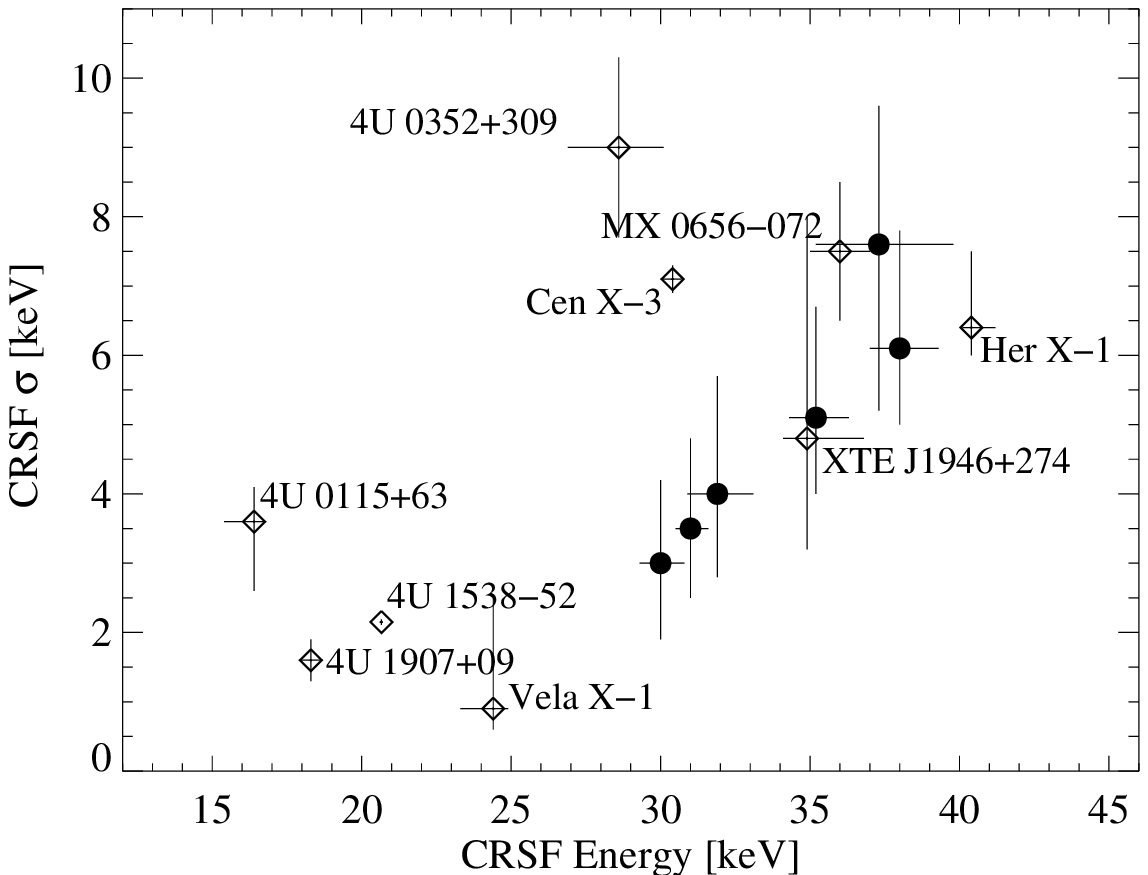}}
\vspace{1ex}
\centerline{\includegraphics[width=0.9\textwidth]{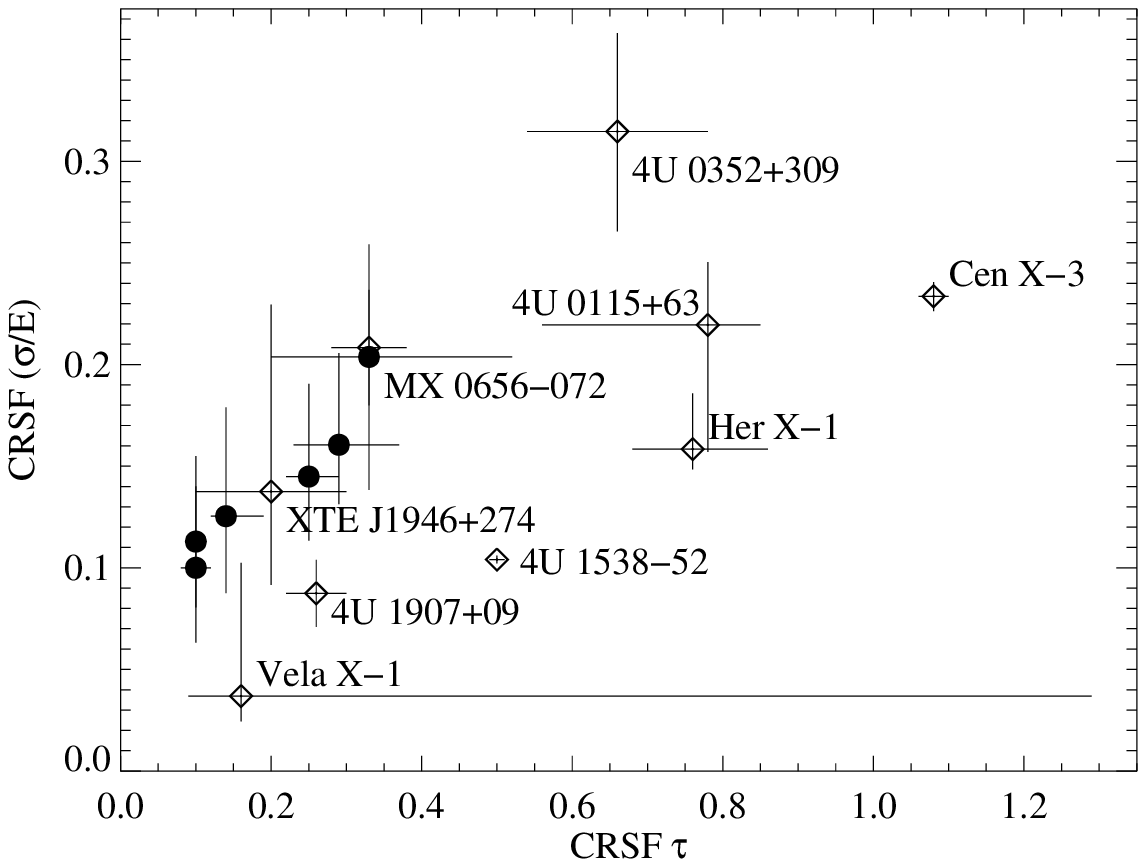}}
\end{minipage}
\caption{\label{fig:gxcorrs} Same as Figs.~\ref{fig:sigmavse} and
\ref{fig:ratiovsdepth}, but with the \gx\ \emph{phase resolved} fits
plotted (solid circles).}
\end{figure}

Unlike \vela, no cyclotron line is detected near the half energy of
the strong deep line at \aprx35\,keV.  This leaves \gx\ a poor fit to
the cutoff energy vs. cyclotron energy correlation
(figure~\ref{fig:hecutvse}).  However, the phase resolved points are a
good fit to the other correlations.  Figure~\ref{fig:gxcorrs} shows
both the width--energy correlation and the relative width--depth
correlation.  The fact that these correlations are confirmed within a
single object is good evidence that they are a physical result of the
cyclotron line formation.  Furthermore, if indeed the electron temperature
is responsible for the line width, then we can infer that the range of
viewing angles to the magnetic field is limited, varying only by
roughly 12\degree\ as the star rotates \cite{Kreykenbohm2004}.

\section{Considerations for the Next X-ray Timing Mission}

Unlike most of the anticipated targets of a next generation timing
mission, we do not expect the accreting pulsars to benefit greatly
from the ability to measure power spectra to very high (\gtsim kHz)
frequencies.  With the exception of photon bubble oscillations
\cite{Klein1996}, these systems are not predicted to have
power spectral features in the kHz range.  This stems from their large
magnetospheric (\aprx 10$^{10}$\,cm) radii where the accretion disk is
truncated and the Keplerian period is of order the neutron star spin
period.  Minimum timescales are therefore limited to roughly the
neutron star spin period, \aprx 0.1--$10^4$\,s.  

Instead, because their spectra can be time variable and are often
strong functions of rotation- (or equivalently, {pulse-)} phase, the
ability to measure the broad band spectrum (\aprx 0.1--100\,keV) with
high significance in a short integration time is the key to future
understanding the emission mechanism and cyclotron line formation. For
example, if the phase-average spectrum is stable over
\aprx10\,ks but the spectrum varies in as little at 0.1 in rotation
phase, then it is necessary to measure the spectrum in only 1\,ks.
Large area detectors are therefore not only required for observing
high frequency timing features, but are also vital for the study of
any source whose spectrum is variable over short timescales.  Finally,
in order to achieve balanced statistics over the steeply falling
pulsar spectra, instrumental effective areas should be heavily
weighted toward the high energies.  This is particularly important
given that all of the known cyclotron lines lie above 10\,keV.




\bibliographystyle{aipproc}   

\bibliography{heindlw}


\end{document}